\documentclass[%
reprint,
superscriptaddress,
 amsmath,
 amssymb,
 aps,
pra,
]{revtex4-2}
\usepackage{graphicx}
\usepackage{subfig}
\usepackage{dcolumn}
\usepackage{bm}
\usepackage{caption}
\usepackage{balance}
\usepackage{bbm}
\usepackage{hyperref}
\hypersetup{
    colorlinks=true,
    linkcolor=red,
    filecolor=red,
    urlcolor=red,
    citecolor=blue, 
}

\usepackage[font=small,labelfont=bf]{caption}
\usepackage[symbol]{footmisc}
\usepackage{stackengine}

\newcommand{\bra}[1]{\left\langle #1 \right|}
\newcommand{\ket}[1]{\left|#1\right\rangle}

\newcommand{\braz}[1]{\langle #1 |}
\newcommand{\ketz}[1]{|#1\rangle}
\newcommand{\braketz}[2]{\langle#1 |  #2\rangle}

\newcommand{\Tr}[1]{\operatorname{Tr}#1}
\makeatletter
\newcommand*{\rom}[1]{\expandafter\@slowromancap\romannumeral #1@}
\makeatother

\let\f\frac
\let\p\partial

\begin{document}


\title{Probabilistic Unitary Formulation of Open Quantum System Dynamics}

\author{\small Le Hu}
\email{le.hu@northwestern.edu}
\affiliation{Department of Physics and Astronomy, University of Rochester, Rochester, New York 14627, USA}
\affiliation{Institute for Quantum Studies, Chapman University, 1 University Drive, Orange, CA 92866, USA}
\affiliation{Department of Physics and Astronomy, Northwestern University, Evanston, Illnois 60208, USA}
\author{\small and Andrew N. Jordan}

\affiliation{Institute for Quantum Studies, Chapman University, 1 University Drive, Orange, CA 92866, USA}
\affiliation{The Kennedy Chair in Physics, Chapman University, Orange, CA 92866, USA}
\affiliation{Department of Physics and Astronomy, University of Rochester, Rochester, New York 14627, USA}
\date{\today}

\begin{abstract}
We show that all non-relativistic quantum processes, whether open or closed, are either unitary or probabilistic unitary, i.e., probabilistic combination of unitary evolutions. This means that for open quantum systems, its continuous dynamics can always be described by the Lindblad master equation with all jump operators being unitary. We call this formalism the probabilistic unitary formulation of open quantum system dynamics. This formalism is shown to be exact under all cases, and does not rely on any assumptions other than the continuity and differentiability of the density matrix. Moreover, it requires as few as $d-1$ jump operators, instead of $d^2-1$, to describe the open dynamics in the most general case, where $d$ is the dimension of Hilbert space of the system. Importantly, different from the conventional Lindblad master equation, this formalism is state-dependent, meaning that the Hamiltonian, jump operators, and rates, in general all depend on the current state of the density matrix. Hence one needs to know the explicit expression of the density matrix in order to write down the probabilistic unitary master equation explicitly. Experimentally, the formalism provides a scheme to control a quantum state to evolve along designed non-unitary quantum trajectories, and can be potentially useful in quantum computing and quantum control scenes since only unitary resources are needed for implementation. \end{abstract}
\maketitle


\section{\label{sec:level1}Introduction}
While the evolution of a closed quantum system can be well-described by the Schr\"{o}dinger equation, realistic quantum systems are almost always affected by their surroundings.  The dynamics of such open quantum system is generally non-unitary and hence requires a different description. Of the numerous descriptions \cite{landau1927dampfungsproblem, lamb1964theory, redfield1965theory, davies1974markovian, KOSSAKOWSKI1972247, gorini1976completely, lindblad1976generators} developed, the best-known one is the master equation in standard (Lindblad) form \cite{KOSSAKOWSKI1972247, gorini1976completely, lindblad1976generators},
\begin{equation}\label{eq1}
	\dot{\rho}=-i[H, \rho]+\sum_{i=1}^{d^{2}-1} \gamma_{i}\left(L_{i} \rho L_{i}^{\dagger}-\frac{1}{2}\left\{L_{i}^{\dagger} L_{i}, \rho\right\}\right).
\end{equation} 
The derivation and applications of the Lindblad master equation often rely on assumptions such as weak coupling limit or semigroup property \cite{breuer2002theory}, and hence are restricted in the Markovian regime. Many efforts have been made to extend the description in the non-Markovian regime, including the Nakajima-Zwanzig equation \cite{nakajima1958quantum, zwanzig1960ensemble}, time-convolutionless projection-operator technique \cite{kubo1963stochastic, chaturvedi1979time}, correlated projection superoperator technique \cite{PhysRevA.74.053815, PhysRevA.75.022103}, collisional models \cite{rybar2012simulation, ciccarello2013quantum, PhysRevA.87.040103}, stochastic Schr\"{o}dinger equation method \cite{imamog1994stochastic, diosi1997non,PhysRevA.58.1699}, rate operator quantum jump technique \cite{PhysRevLett.124.190402} and many more \cite{RevModPhys.89.015001}. Nevertheless, establishing a simple, intuitive, and unified framework to describe the dynamics of open quantum systems in all regimes is still a long-standing open problem.

In this paper, we are going to show, by deriving merely from an ansatz, that the dynamics of any finite, $d$-dimensional (later generalized to the infinite-dimensional case), continuously evolving open quantum system $\rho(t)$ can be described exactly by a time- and state-dependent Hamiltonian and probabilistic combinations of $d-1$, instead of $d^2-1$ (as in Eq.\,(\ref{eq1})), time- and state-dependent unitary operators,
\begin{equation} \label{eq2}
	\dot{\rho}(t)=-i[H(t),\rho(t)]+\sum_{i=0}^{d-1}q_i(t)(\tilde{\mathcal{U}}_{i,t} \rho(t)\tilde{\mathcal{U}}^\dagger_{i,t}-\rho(t)),
\end{equation}
where $H(t)$ is the Hamiltonian that can drive all instantaneous eigenvectors $\{ \ket{\psi_i(t)}\}$ of the density matrix $\rho(t)$ of the system, i.e.,
\begin{equation}\label{eq3}
	H(t)\ket{\psi_i(t)}=i\ket{\p_t\psi_i(t)} \quad i=1,2, \dots ,d,
\end{equation}
$\tilde{\mathcal{U}}_{i,t}$ is the $i$th traceless (except for $i=0$), linearly-independent (i.e., $\operatorname{Tr}[\tilde{\mathcal{U}}_{i,t}^\dagger \tilde{\mathcal{U}}_{j,t}]=\delta_{ij}d$) and time-dependent unitary operators defined by
\begin{equation}
\tilde{\mathcal{U}}_{i,t}\equiv \mathbb{U}_t \mathcal{U}_{i}\mathbb{U}^\dagger_t,
\end{equation}
where $\mathbb{U}_t$ is the unitary operator that diagonalizes the density matrix $\rho(t)$,
\begin{equation}\label{eq4}
\rho(t)=\mathbb{U}_t\rho_D(t)\mathbb{U}^\dagger_t,\end{equation}
and $\mathcal{U}_{i} \in\{U_{n m} \in \mathbb{R}^{d \times d} \mid n, m \in\{0,1, \ldots, d-1\}\}$ is the $i$th real Weyl operator \cite{bertlmann2008bloch} (especially, $\mathcal{U}_0=\tilde{\mathcal{U}}_{0,t}=\mathbbm{1}$),
\begin{equation}\label{eq5}
	U_{nm}=\sum_{k=0}^{d-1} \mathrm{e}^{\frac{2 \pi \mathrm{i}}{d} k n}|k\rangle\langle(k+m) \bmod d|.
\end{equation}
The $q_i(t)$,  which is solvable from a system of equations, denotes the rate so that $q_i(t)dt$ is the probability that $\tilde{\mathcal{U}}_{i,t}$ is applied to the quantum state during the time interval $[t, t+dt]$. Since $q_i(t)dt$ has a clear physical meaning, one can control a quantum state to evolve along a designed quantum trajectory $\rho(t)$, by setting a short time interval $dt$ and applying a time-dependent Hamiltonian $H(t)$ and unitary operators $\tilde{\mathcal{U}}_{i,t}$ with classical probability $q_i(t)dt$ during the time interval $[t,t+dt]$ (see Appendix \ref{appendix:a} for a numerical implementation). For instance, such applications of unitary operators can be randomly determined by a classical computer such that for ignorant observers, the quantum states appear to evolve non-unitarily to them based on their measurements \cite{peetz2023simulation}. It has to be noted that we can go beyond the description given above by letting $q_i(t)$ be negative \cite{feynman1984negative} or even singular for finite-dimensional systems, the significance of which we shall discuss in detail. Since experimentally one cannot make an event happen with negative or singular probability, though such phenomena occur naturally as we shall see, such a quantum control scheme will only work in the cases where $q_i(t) \geq 0$ for all $i$, which corresponds to the evolution with contracting trace distance \cite{nielsen2002quantum} and non-increasing purity (Appendix \ref{appendix:b}). Hence when such a quantum control scheme is not applicable, other means of control \cite{alipour2020shortcuts, hu2020quantum,PhysRevLett.127.270503, suri2023two} should be applied.

\section{\label{sec:level2}Probabilistic unitary formulation}
Below we will derive in detail the formulation for the finite-dimensional case, which is completely based on the ansatz that the continuous dynamics of any finite-dimensional open quantum system can be regarded as the combined effects of a probabilistic combination of unitary operators, plus a time-dependent Hamiltonian. Mathematically, the ansatz can be formulated as
\begin{equation}\label{eq6}
\begin{aligned}
	\rho(t+dt)&=(1-\sum_{i=1}^mq_i(t)dt)U_t(dt)\rho(t)U^\dagger_t(dt)\\&+\sum_{i=1}^m(q_i(t)dt\, \tilde{\mathcal{U}}_{i,t}\rho(t)\tilde{\mathcal{U}}^\dagger_{i,t}),
	\end{aligned}
\end{equation}
where $U_t(dt)$ and $\tilde{\mathcal{U}}_{i,t}$ are some unitary operators yet to be determined. The physical meaning of the equation is that during the time interval $[t,t+dt]$, there is probability $q_i(t)dt$ that the quantum state will be subject to the unitary transformation $\tilde{\mathcal{U}}_{i,t}$, and probability $1-\sum_{i=1}^mq_i(t)dt$ that the quantum state will be subject to a time-dependent Hamiltonian, characterized by $U_t(dt)\equiv \exp(i H(t)dt)$. By Taylor expanding $U_t(dt)=(1-i H(t)dt+\mathcal{O}(dt^2))$, and calculating $\dot{\rho}(t)$ according to its limit definition $\dot{\rho}(t)=\lim_{dt\to0} (\rho(t+dt)-\rho(t))/dt$, one can straightforwardly recover Eq.\,(\ref{eq2}), except that the minimum value of $m$ for a $d$-dimensional Hilbert space, and the explicit expression of $H(t)$, $\tilde{\mathcal{U}}_{i,t}$, and $q_i(t)$, are undetermined.

To determine their explicit expressions, let us first consider the simpler case that the density matrices at time $t$ and $t+dt$ happen to be simultaneously diagonalized, denoted as $\rho_D(t)$ and $\rho_D(t+dt)$, and that $H(t)=0$. In this simplified scenario, only the eigenvalues of the density matrix change over time, hence it becomes much easier to track their changes. Let us denote the eigenvalues of $\rho_D(t)$ by $p_1(t), \dots,p_d(t)$, and the eigenvalues of $\rho_D(t+dt)$ by $p_1(t)+f_1(t)dt, \dots, p_d(t)+f_d(t)dt$; we need to find a set of $\{\mathcal{U}_{i,t}\}$ to describe the changes of eigenvalues via Eq.\,(\ref{eq6}). It turns out that $\{\mathcal{U}_i\}$ can be chosen to be the real Weyl operators (Eq.\,(\ref{eq5})), where the $i$ denotes the number of 1's in the lower left conner of $\mathcal{U}_i$. The effect of those real Weyl operators is that they ``rotate'' the eigenvalues cyclically, such that $\operatorname{diag}(\mathcal{U}_i\rho_D(t)\mathcal{U}^\dagger_i)=(p_{1+i}, p_{2+i}, \dots, p_{d+i})$, where the $+$ here is defined as modular addition (i.e., $i+j \to (i+j) \mod d$). There are $d-1$ non-identity real Weyl operators, and since we have assumed $H(t)=0$, we have $U_t(dt)=\mathbbm{1}$, resulting in a total number of $d$ operators, a minimum sufficient number to cycle the eigenvalues  back. Plugging in the expression of $\mathcal{U}_{i}$, $U_t(dt)$, $\rho_D(t)$, and $\rho_D(t+dt)$ into Eq.\,(\ref{eq6}), one obtains a system of equations, which can be written in the matrix form
\begin{equation} \label{eq7}
\underbrace{\begin{pmatrix}
	p_1&p_d&p_{d-1}&\cdots& p_2\\
	p_2&p_1&p_d &\cdots & p_3\\
	\vdots &&\cdots &&\vdots\\
	p_d &p_{d-1} &p_{d-2}&\cdots&p_1
\end{pmatrix}}_{P}
\underbrace{
\begin{pmatrix}
	-q_{0}\\
	q_{1}\\
	\vdots\\
	q_{d-1}
\end{pmatrix}}_{\vec{q}}=\underbrace{\begin{pmatrix}
	f_{1}\\
	f_{2}\\
	\vdots\\
	f_{d}
\end{pmatrix}}_{\vec{f}}
\end{equation}
where we have defined $q_0(t)=\sum_{i=1}^{d-1}q_i(t)$. The matrix $P$, which is \textit{doubly stochastic} (i.e., $\sum_i P_{ij}=\sum_j P_{ij}=1$ where $P_{ij} \geq 0$) and \textit{circulant} (i.e., a square matrix where each row can be generated by rotating cyclically the elements of its preceding row), can be easily diagonalized by the discrete Fourier transform and is almost always non-singular, so that we have the unique solution for $\vec{q}$ for a given $\vec{f}$. Moreover, it can be shown by straightforward algebra that $\sum_{i=1}^d f_i=0$ if and only if $-q_0+\sum_{i=1}^{d-1} q_i=0$, indicating that the solution $\vec{q}$ conserves probability if and only if $\vec{f}$ conserves the trace. To address the possible singularity of $P$, we have the following theorem:

Theorem \cite{kra2012circulant}. \textit{If $\{v_j\}_{0\leq j \leq {n-1}}$ is a nonincreasing or nondecreasing sequence of nonnegative or nonpositive real numbers, then the circulant matrix $V = \mathrm{circ}{(v_0, v_1, \cdots, v_{n-1})}$ is singular if and only if for some integer $d\mid n$, $d \geq 2$, the vector $v=(v_0,v_1,...,v_{n-1})$ consists of $\f{n}{d}$ consecutive constant blocks of length $d$. }

Since the eigenvalues of the density matrix are always nonnegative, and since we can always rearrange the eigenvalues in a nonincreasing or nondecreasing manner by a unitary transformation, the above theorem is applicable in our case, meaning that $P$ is singular if and only if the eigenvalues of the density matrix can be arranged into consecutive constant blocks of the same length (e.g., $\operatorname{diag}(\rho_D)=(a,a,b,b,c,c)$ where $a\geq b\geq c$). When such singular $P$ indeed occurs, one can calculate $P^{-1}$ analytically in the regime where $P$ is invertible, and assume $P$ is invertible everywhere to obtain the solution of $\vec{q}$. Such solution of $\vec{q}$ will be singular where $P^{-1}$ is singular, yet we stress that one can still use the solution of $\vec{q}$ to describe the dynamics. Such rate $\vec{q}$ with singular point(s) occurs naturally, for example, when a qubit goes through its maximally mixed state, as we shall see in Example I. It is worth noting that while such singularity does not hamper the description of the dynamics, it will make the quantum control scheme break down \textit{around} the singular points of $\vec{q}$, as $q_i(t)dt$ for some $i$ will blow up and become unphysical for a preset time interval $dt$.

We have shown how Eq.\,(\ref{eq2}) can be used to describe the open quantum system dynamics when the density matrix happens to be diagonalized and $H(t)=0$. One can then use Eq.\,(\ref{eq6}) and the limit definition $\dot{\rho}_D(t)=\lim_{dt \to 0} (\rho_D(t+dt)-\rho_D(t))/dt$ to obtain the \textit{master equation for the diagonalized density matrix},
\begin{equation} \label{eq8}
	\dot{\rho}_D(t)=-q_0(t)\rho_D(t)+\sum_{i=1}^{d-1}q_i(t) \mathcal{U}_i \rho_D(t)\mathcal{U}^\dagger_i,
\end{equation}
where the rate $q_i(t)$ is solvable from Eq.\,(\ref{eq7}) and $\mathcal{U}_i$ is the $i$th real Weyl operators.

To generalize the above description to the non-diagonal cases, one only needs to establish explicitly the relation between $\rho_D(t)$, $\dot{\rho}_D(t)$ and $\rho(t)$, $\dot{\rho}(t)$, respectively, and plug the expression of $\rho_D(t)$ in terms of $\rho(t)$, and  $\dot{\rho}_D(t)$ in terms of $\dot{\rho}(t)$, into the above equation. To do so, assume that $\rho(t)$ can be diagonalized by $\mathbb{U}_t$, i.e., $\rho(t)=\mathbb{U}_t \rho_D(t)\mathbb{U}^\dagger_t$, and that $\rho(t+dt)$ can be diagonalized by $\mathbb{U}_{t+dt}$. Since $\rho(t)$ is assumed to evolve continuously, $\mathbb{U}_{t+dt}$ should only differ from $\mathbb{U}_{t}$ by an infinitesimal amount. In other words, there should exist a unitary operator, denoted as $U_t(dt)$, which can connect $\mathbb{U}_{t}$ and $\mathbb{U}_{t+dt}$, such that $\mathbb{U}_{t+dt}=U_t(dt)\mathbb{U}_t$. Since the column vectors of $\mathbb{U}_t$ and $\mathbb{U}_{t+dt}$ are composed by the instantaneous eigenvectors $\ket{\psi_i(t)}$ and $\ket{\psi_i(t+dt)}$ of $\rho(t)$ and $\rho(t+dt)$, respectively, the requirement of $\mathbb{U}_{t+dt}=U_t(dt)\mathbb{U}_t$ is equivalent to the requirement of $U_t(dt)\ket{\psi_i(t)}=\ket{\psi_i(t+dt)}$, for each $i=1,2,\dots,d$. This is just the Schr\"{o}dinger equation should we expand $U_t(dt)=(1-iH(t)dt+\mathcal{O}(dt^2))$ and also $\ket{\psi_i(t+dt)}$ by Taylor series, and the Hamiltonian $H(t)$ has to be such that it solves $\ket{\psi_i(t)}$ for all $i$. A scheme to find the Hamiltonian has been solved in our recent work \cite{hu2022quantum}, where one can construct the Hamiltonian, which is also state-dependent, by
\begin{equation} \label{eq10}
	 H(t)=i \sum_{i=1}^d\ketz{\partial_t \tilde{\psi}_i(t)}\braz{\tilde{\psi}_i(t)},
\end{equation}
where  $\ketz{\tilde{\psi}_i(t)}\equiv e^{i\phi_i(t)}\ket{\psi_i(t)}$ and $\phi_i(t)\equiv\int-i\left\langle\partial_{t} \psi_i(t)|\psi_i(t)\right\rangle d t$. The Hamiltonian constructed in this way is optimal in the sense that it has the minimum Hilbert-Schmidt norm $\| H\|_{HS}=\operatorname{Tr}[H^2(t)]$ (Appendix \ref{appendix:c}), making its experimental implementation easier. One may also simply let $H(t)=i \sum_i^d \ketz{\partial_t {\psi}_i(t)}\braz{{\psi}_i(t)}$ if such optimization is unneeded.

To establish the relation between $\dot{\rho}(t)$ and $\dot{\rho}_D(t)$, first write $\dot{\rho}(t)$ in its limit definition $\dot{\rho}(t)=\lim_{dt \to 0}(\rho(t+dt)-\rho(t))/dt$, then plug in $\rho(t)=\mathbb{U}_t \rho_D(t)\mathbb{U}^\dagger_t$ and $\rho(t+dt)=\mathbb{U}_{t+dt} \rho_D(t+dt)\mathbb{U}^\dagger_{t+dt}$, and finally make use of $\mathbb{U}_{t+dt}=U_{t}(dt)\mathbb{U}_t$, $U_t(dt)=(1-iH(t)dt+\mathcal{O}(dt^2))$ and $\rho_D(t+dt)=\rho_D(t)+\dot{\rho}_D(t)dt+\mathcal{O}(dt^2)$, from which  one obtains
\begin{equation}
	\dot{\rho}(t)=-i[H(t), \rho(t)]
+\mathbb{U}_t \dot{\rho}_{D}(t) \mathbb{U}_t^\dagger,\end{equation}
or equivalently,
\begin{equation}
	\dot{\rho}_{D}(t)=\mathbb{U}_t^{\dagger}(\dot{\rho}(t)+i[H(t), \rho(t)]) \mathbb{U}_t.
\end{equation}
By plugging the above equation of $\dot{\rho}_D(t)$ and $\rho_D(t)=\mathbb{U}_t^\dagger \rho(t) \mathbb{U}_t$ into Eq.\,(\ref{eq8}), one can readily recover the main result of the paper, Eq.\,(\ref{eq2}). 

Since the only assumption made in the above derivations is the continuity of $\rho(t)$, Eq.\,(\ref{eq2}) is exact under all cases and thus very general. Moreover, from  Eq.\,(\ref{eq2}), one can immediately obtain the semigroup master equation, by taking the generator $\mathcal{L}_t$, defined by $\dot{\rho}(t)=\mathcal{L}_t \rho(t)$, to be time-independent. A simple choice then is letting $H(t)$, $\tilde{\mathcal{U}}_{i,t}$ and $q_i(t)$ all to be time-independent, and the complete positivity further requires $q_i\geq0$ for all $i$ by the Gorini-Kossakowski-Sudarshan-Lindblad theorem \cite{gorini1976completely, lindblad1976generators}. Note that the time independency of $H$, $\tilde{\mathcal{U}}_{i}$ and $q_i$ is only a sufficient requirement, not a necessary one. As we shall see in Example II, contrary to the usual belief, a semigroup master equation can also contain time-dependent parameters, if it is written in the state-dependent form.

\section{\label{sec:level1}Discussions}
We highlight that the formalism developed above can in fact be used to describe the master equation of any Hermitian matrix $\mathcal{H}(t)$ via Eq.\,(\ref{eq2}). In this general case, since $\vec{f}$ does not have to conserve the trace, the solution $\vec{q}$ does not have to conserve the probability. On the other hand, one can also use the formalism to describe the generalized quantum channel that may not be complete positive or even not positive, which could occur naturally when the initial state is correlated with the environment \cite{PhysRevLett.73.1060,shaji2005s}. If we denote the dynamical map by $\Phi_t(s)$, then \textit{any} quantum channel $\rho(t+s)=\Phi_t(s)[\rho(t)]$ can be described by (Appendix \ref{appendix:d})
\begin{equation}\label{eq12}
\begin{aligned}
	\rho(t+s)&=\sum_{i=0}^{d-1} q_i(t,s) \tilde{\mathcal{U}}_{i;t,s} \rho(t) \tilde{\mathcal{U}}^\dagger_{i;t,s},
	\end{aligned}
\end{equation}
where
\begin{equation}
	\tilde{\mathcal{U}}_{i;t,s} =U_t(s)\mathbb{U}_t\mathcal{U}_i \mathbb{U}^\dagger_t.
\end{equation}
The $U_t(s)$ is defined by $\mathbb{U}_{t+s}=U_t(s)\mathbb{U}_t$, where $\mathbb{U}_{t+s}$ and $\mathbb{U}_t$ are the unitary matrices diagonalizing $\rho(t+s)$ and $\rho(t)$, respectively. The $q_i(t,s)$, which can be solved by a similar manner from Eq.\,(\ref{eq7}) (especially, $q_0(t,s)\equiv 1-\sum_{i=1}^{d-1}q_i(t,s)$), denotes the (possibly negative or singular) probability the $\tilde{\mathcal{U}}_{i;t,s}$ is applied to $\rho(t)$. If $\sum_i q_i(t,s)=1$ and $q_i(t,s) \in [0,1]$ for all $i$, then the quantum channel $\Phi_t(s)$ is called \textit{mixed unitary} or \textit{random unitary} \cite{audenaert2008random, watrous2018theory}. It is interesting to note that Lee and Watrous \cite{lee2020detecting} proved that detecting whether a quantum channel is mixed unitary is NP-hard, yet we just show that any quantum channel can be written in the mixed unitary form if one relaxes the restrictions on $q_i(t,s)$ by allowing negative and singular probability. Moreover, one can verify that
\begin{equation}\label{eq14}
	\sum_{i=0}^{d-1}\underbrace{(\sqrt{q_i(t,s)}\tilde{\mathcal{U}}_{i;t,s})}_{\mathcal{K}_i} \underbrace{(\sqrt{q_i(t,s)}\tilde{\mathcal{U}}^\dagger_{i;t,s})}_{\bar{\mathcal{K}}_i}=\mathbbm{1},
\end{equation} and rewrite Eq.\,(\ref{eq12}) by
\begin{equation} \label{0eq15}
	\rho(t+s)=\sum_{i=0}^{d-1} \mathcal{K}_i \rho(t)\bar{\mathcal{K}}_i,
\end{equation}
which has a very similar form as the Kraus representation $\sum_i K_i \rho(t)K_i^\dagger$ with the constraint $\sum_i K_i K_i^\dagger=\mathbbm{1}$, except that in the above case, $\bar{\mathcal{K}}_i=\mathcal{K}_i^\dagger$ (if $q_i(t,s) \geq 0$) or $\bar{\mathcal{K}}_i=-\mathcal{K}_i^\dagger$ (if $q_i(t,s) \leq 0$). Different from the Kraus representation, which works only for the trace-preserving and complete positive map, Eq.\,(\ref{0eq15}) under the constraint Eq.\,(\ref{eq14}) is valid for any trace-preserving map. Note that just as in the continuous case, Eq.\,(\ref{0eq15}) is also state-dependent.

The whole formalism above can be easily generalized to the countably infinite-dimensional case, which delivers the master equation
\begin{equation} \label{eq17}
	\dot{\rho}(t)=-i[H(t),\rho(t)]+\sum_{i=0}^{\infty}q_i(t)(\tilde{\mathcal{U}}_{i,t} \rho(t)\tilde{\mathcal{U}}^\dagger_{i,t}-\rho(t)),
\end{equation}
where $H(t)$ and $\tilde{\mathcal{U}}_{i,t}$ are defined the same way as in Eq.\,(\ref{eq3})-(\ref{eq4}), and $\mathcal{U}_i$ is instead defined as
\begin{equation}
	\mathcal{U}_i=\sum_{m=0}^\infty\ket{m+i}\bra{m}, \quad i\in\mathbb{N},
\end{equation}
which satisfies the properties $\Tr{(\mathcal{U}^\dagger_i\mathcal{U}_j)}=\Tr{(\tilde{\mathcal{U}}^\dagger_{i,t}\tilde{\mathcal{U}}_{j,t})}=\sum_{k=1}^\infty \delta_{ij}$. The detailed derivation, shown in the Appendix \ref{appendix:e}, shows that for the infinite-dimensional case, $\vec{q}$ is always nonsingular as long as $\vec{f}$ is nonsingular. This is because the $P$ matrix, which would be defined by $P_{ij}=p_{i-j+1}$, where $p_0=p_{-1}=p_{-2}=\dots=0$ and $p_1, p_2, \dots,$ are the eigenvalues of $\rho(t)$, turns out to be an \textit{infinite-dimensional lower triangular Toeplitz matrix} and can always be made invertible by choosing $p_1\neq0$.


Below we will show a few examples followed with comments to demonstrate our formalism.
\section{Examples}
\subsection{Example I - The Jaynes-Cummings model.}
Consider the Jaynes-Cummings model \cite{jaynes1963comparison} under the rotating-wave approximation:
\begin{equation}
	H_{SE}=\hbar \omega_{c} a^{\dagger} a+\hbar \omega_{a} \frac{\sigma_{z}}{2}+\frac{\hbar \Omega}{2}\left(a \sigma_{+}+a^{\dagger} \sigma_{-}\right),
\end{equation}
where $\sigma_\pm=\sigma_x \pm i\sigma_y$. The model describes a two-level atom of resonant frequency $\omega_a$ interacting with a single-mode field in a cavity with field frequency $\omega_c$ and interaction strength $\Omega$. Assuming that initially, the cavity is in the vacuum state $\ket{0}$ and the atom is in the excited state $\ket{e}$ with $\omega_a=\omega_c$, then the state of the total system at a later time is given by \cite{berman2011principles}
\begin{equation}
	|\psi(t)\rangle=\cos \left(\frac{\Omega t}{2}\right)|e, 0\rangle-i \sin \left(\frac{\Omega t}{2}\right)|g, 1\rangle.
\end{equation}
We can then write the above solution in the density matrix form $\rho(t)=\ket{\psi(t)}\bra{\psi(t)}$ and take the partial trace over the cavity to obtain the density matrix of the atom $\rho_S(t)$,
\begin{equation}\label{eq15}
	\rho_S(t)=\operatorname{Tr}_E(\rho)=\left(\begin{array}{cc}\cos ^{2}\left(\frac{\Omega t}{2}\right) & 0 \\ 0 & \sin ^{2}\left(\frac{\Omega t}{2}\right)\end{array}\right),
\end{equation}
which describes the evolution of the atom alone by ignoring the state of the photon. We would like to write down a master equation for $\rho_S(t)$ by using the formulation we developed, which can be obtained by the direct application of Eq.\,(\ref{eq8}) since $\rho_S(t)$ is of the diagonal form,
\begin{equation} \label{eq16}
	\dot{\rho}_S(t)=-i[H_S, \rho(t)]+q_1(t)(\sigma_x \rho_S(t) \sigma_x-\rho_S(t)),
\end{equation}
where $q_1(t)=\frac{1}{2}\Omega \tan({\Omega t})$.

There are a few comments worth mentioning regarding the solution. First, in this specific example, the Hamiltonian $H_S$ can be taken to be anything as long as $[H_S, \rho(t)]=0$, which is also evident from Eq.\,(\ref{eq8}). 
Second, since $q_1(t)$ denotes the rate of an event happening, it is not unphysical at all if $q_1(t) \to \infty$ as $t \to \frac{\pi}{2 \Omega}$. To confirm this, we can calculate $\dot{\rho}(t)$ at time $t=\frac{\pi}{2 \Omega}$ via Eq.\,(\ref{eq16}) to obtain $\dot{\rho}_S(t)=\operatorname{diag}(-\f{\Omega}{2}, \f{\Omega}{2})$, which is exactly what we would obtain via Eq.\,(\ref{eq15}). Interestingly, this is not the case for the mixed unitary channel, which is unital, i.e., $\Phi(\mathbbm{1})=\mathbbm{1}$, meaning that a maximally mixed state will always remain maximally mixed after the application of a unital channel, hence $\dot{\rho}_S=0$. Yet we see clearly that the $\rho_S(t)$ above escapes from being maximally mixed at the time $t=\f{\pi}{2\Omega}$ and $\dot{\rho}_S(\f{\pi}{2 \Omega})\neq0$. This phenomenon, which is counterintuitive and breaks the property of the unital channel, is related with the singular probability. In the above example, even though the two matrices, $-q_1(t)\rho_S(t)$ and $q_1(t)\sigma_x\rho_S(t)\sigma_x$, become singular at $t=\f{\pi}{2 \Omega}$, their sum is nevertheless finite because the infinite terms cancel with each other, somewhat resembling the renormalization in the quantum field theory. It is worth noting that such singular rate, which has been previously reported in literature \cite{chruscinski2010non}, contains rich physics and it is related with entanglement sudden death \cite{yu2004finite, yu2006sudden}. By Eq.\,(\ref{eq7}), the singular rate occurs whenever the matrix $P$ is non-invertible, which depends solely on the eigenvalues of the density matrix.

\subsection{Example II - Decay of a two-level atom.}
Consider the well-known model of a two-level atom spontaneously decaying due to its interaction with the vacuum, the process of which can be described by
\begin{equation} \label{eq23}
	\dot{ \rho}(t)=-i[H, \rho(t)]+\Gamma\left(\sigma^{-} \rho(t) \sigma^{+}-\frac{1}{2}\left\{\sigma^{+} \sigma^{-}, \rho(t)\right\}\right),
\end{equation}
where $\Gamma$ denotes the coupling strength between the atom and the vacuum and $H=\hbar \omega \sigma_z/2$. For simplicity, let us assume that the atom is initially at the excited state. Then by straightforward calculation, it can be shown that the dynamics can be equivalently described by
\begin{equation} \label{eq24}
	\dot{ \rho}(t)=-i[H, \rho(t)]+\f{\Gamma \rho_{11}(t)}{\rho_{11}(t)-\rho_{22}(t)}(\sigma_x \rho(t) \sigma_x-\rho(t)).
\end{equation}
Notice how Eq.\,(\ref{eq23}), a semigroup master equation, can actually contain time-dependent (and even negative) parameters when it is cast into the probabilistic unitary form Eq.\,(\ref{eq24}). Different from usual master equations such as Eq.\,(\ref{eq1}), the probabilistic unitary master equation generally is state-dependent, meaning that if we have a different initial state, then the Eq.\,(\ref{eq24}) would have a different form. Hence one must write down different probabilistic unitary master equations for different quantum trajectories even though such trajectories can be described by a single master equation by the conventional method (see Appendix \ref{appendix:f} for an example).

\section{Conclusions}
In this paper, we have derived the master equations which can exactly describe open quantum system dynamics across all regimes. These equations, i.e., Eq.\,(\ref{eq2}), (\ref{eq12}) and (\ref{eq17}), combined with the Schr\"{o}dinger equation, suggest that all non-relativistic quantum processes are either unitary or probabilistic unitary, including both the continuous and discontinuous (i.e., quantum jump) dynamics in an open or closed quantum system with finite or infinite Hilbert space. The results therefore unify quantum-mechanical dynamics by providing a unitary-operator description in the nonunitary regime, enabling us to understand non-unitary quantum processes, such as decoherence and wave-function collapse, in a more coherent perspective. Practically, it also suggests a scheme to control quantum state to evolve along designed trajectories, which could be potentially useful in quantum control schemes as only unitary resources are needed. Finally, it is worth noting that the techniques used in this paper can be applied to show that the signs of the rates in the Lindblad master equation can always be arbitrarily determined, including cases where all rates are always non-negative \cite{hu2023signs}, provided the master equation is state-dependent, as in this paper.

\section{Acknowledgement}
We thank Luiz Davidovich for inspiring discussions. We are also grateful to the support from the Army Research Office (ARO) under the grant W911NF-22-1-0258.

\begin{widetext}
\appendix
 \setcounter{equation}{0}
\setcounter{figure}{0}
\setcounter{table}{0}
\makeatletter
\renewcommand{\theequation}{A\arabic{equation}}
\renewcommand{\thefigure}{A\arabic{figure}}

 \begin{center}
\section{Numerical simulations of quantum control by probabilistic unitary method\label{appendix:a}}
\end{center}
In the following, we show numerical simulation results of the decay of a two-level atom via the probabilistic unitary master equation Eq.\,(\ref{eqs21}). The simulation was run on a classical computer, with initial state $\ket{+} = \frac{1}{\sqrt{2}}(\ket{1}+\ket{0})$. At each time $t_0$, a random real number between 0 and 1 is generated; if the random number is smaller than $q_1(t_0)\Delta t$, then $\tilde{\mathcal{U}}_{t_0}$ (Eq.\,\ref{eqs22}) is applied to the quantum state $\rho(t_0)$. Otherwise, $U(t_0,\Delta t)=\exp (-iH(t_0)\Delta t)$ is applied, where $H(t_0)$ is given by Eq.\,(\ref{eqs23}). Fig.\,(\ref{fig:epsart1})(a) showes the numerical (data points) and analytical (curve) results of $\langle \sigma_x \rangle$, $\langle \sigma_y \rangle$ and $\langle \sigma_z \rangle$ for the setup $\Gamma=0.2, \omega =1, \hbar=1, \Delta t=0.01 \pi$. Each data point represents $10^3$ simulations. Fig.\,(\ref{fig:epsart1})(b) shows the same simulation, but for each time step, there is a 0.1\% chance that an X, Y, or Z error occurs.
\begin{figure*}[!htp]

\subfloat{\includegraphics[width=.5\textwidth,scale=1]{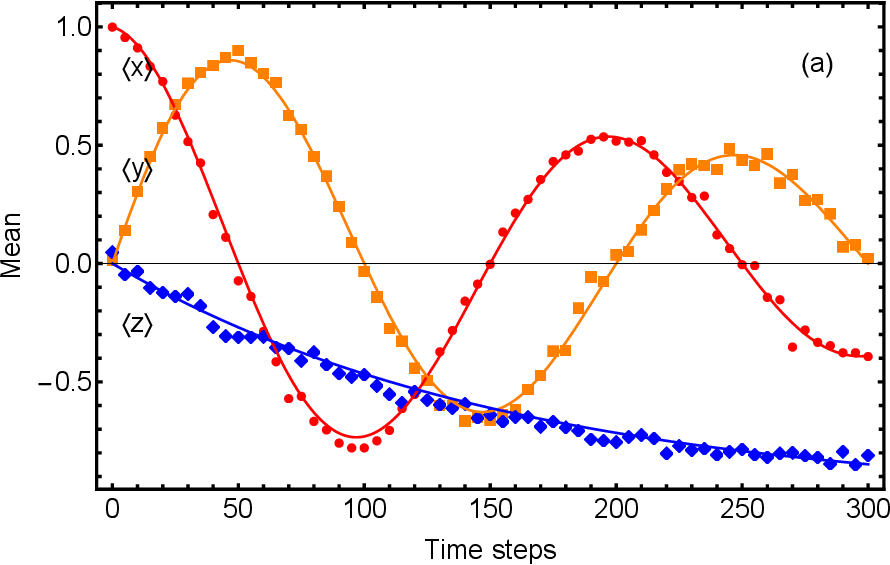}\hfill}
\subfloat{\includegraphics[width=.5\textwidth,scale=1]{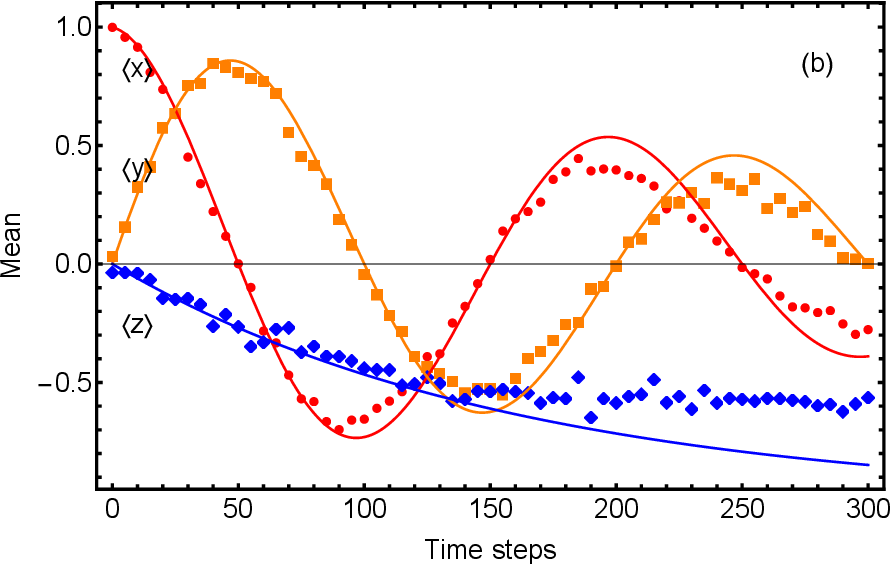}\hfill}
\caption{\label{fig:epsart1} Panel (a) displays numerical (data points) and analytical (curves) results of $\langle \sigma_x \rangle$, $\langle \sigma_y \rangle$ and $\langle \sigma_z \rangle$ of the decay of a two-level atom. Each data point represents $10^3$ simulations run on a classical computer using a probabilistic unitary master equation. Panel (b) shows the same simulation, but for each time step, there is a 0.1\% chance that an X, Y or Z error occurs. In both figures, we take $\Gamma=0.2, \omega =1, \hbar=1$, and each time step lasts $\Delta t=0.01 \pi$.
} 
\end{figure*}

\setcounter{equation}{0}
\setcounter{figure}{0}
\setcounter{table}{0}
\makeatletter
\renewcommand{\theequation}{B\arabic{equation}}
\renewcommand{\thefigure}{B\arabic{figure}}
\bigskip
\section{Proof of non-increasing purity of the probabilistic unitary master equation with non-negative rates\label{appendix:b}}

Below, we will show that if $q_i(t) \geq 0$ for all $i$, then Eq.\,(\ref{eq2}) describes dynamics with non-increasing purity.
Since
	$\f{d}{dt}\operatorname{Tr}[\rho^2(t)]=2\Tr{(\dot{\rho}(t)\rho(t))}$, by plugging $\dot{\rho}(t)$ defined by  Eq.\,(\ref{eq2}) into $2\Tr{(\dot{\rho}(t)\rho(t))}$, one obtains
	\begin{equation}
	\begin{aligned}
		\f{d}{dt}\operatorname{Tr}[\rho^2(t)]&=2 \Tr{\left[\left( -i[H(t),\rho(t)]+\sum_{i=0}^{d-1}q_i(t)(\tilde{\mathcal{U}}_{i,t} \rho(t)\tilde{\mathcal{U}}^\dagger_{i,t}-\rho(t))\right)\rho(t)\right]}\\
		&=\underbrace{-2\Tr(i[H(t),\rho(t)]\rho(t))}_{=0}+2\sum_{i=0}^{d-1}q_i(t)\Tr[\underbrace{\tilde{\mathcal{U}}_{i,t}\rho(t)\tilde{\mathcal{U}}^\dagger_{i,t}}_{\equiv \rho_i^\prime(t)}\rho(t)]-2\sum_{i=0}^{d-1}q_i(t)\Tr[\rho^2(t)]\\
		&=2\sum_{i=0}^{d-1}q_i(t)(\Tr[\rho_i^\prime(t)\rho(t)]-\Tr{[\rho^2(t)]}),
	\end{aligned}
	\end{equation}
	where the term $-2\Tr(i[H(t),\rho(t)]\rho(t))=0$ by the cyclic permutations of the trace. If one decomposes the density matrix into its Bloch vector form $\rho(t)=\f{1}{d}I+\vec{v}_\rho \cdot \vec{\Lambda}$, where $\vec{\Lambda}$ is formed by its SU($n$) basis (e.g., generalized Gell-Mann matrices) \cite{bertlmann2008bloch}, then
	\begin{equation}
		\Tr[\rho_i^\prime(t)\rho(t)]=\vec{v}_{\rho^\prime}\cdot \vec{v}_{\rho} = |\vec{v}_{\rho^\prime}||\vec{v}_{\rho}|\cos{\theta} \leq |\vec{v}_{\rho}||\vec{v}_{\rho}|=\vec{v}_{\rho} \cdot \vec{v}_{\rho}=\Tr{[\rho^2(t)]}
	\end{equation}
	where $\theta$ is the angle between $\vec{v}_\rho$ and $\vec{v}_{\rho^\prime}$. Note that $|\vec{v}_\rho|=|\vec{v}_{\rho^\prime}|$ since unitary transformation does not change the length of the Bloch vector \cite{bertlmann2008bloch}.
	If $q_i(t)\geq0$ for all $i$, one can then readily conclude $\f{d}{dt}\operatorname{Tr}[\rho^2(t)] \leq 0$.

 \setcounter{equation}{0}
\setcounter{figure}{0}
\setcounter{table}{0}
\renewcommand{\theequation}{C\arabic{equation}}
\renewcommand{\thefigure}{C\arabic{figure}}

\section{Proof of the minimum Hilbert-Schmidt-normed Hamiltonian\label{appendix:c}}

To see that the Hamiltonian $H(t)=i \sum_{i} |\partial_{t} \tilde{\psi}_{i}(t) \rangle \langle\tilde{\psi}_{i}(t) |$ defined by Eq.\,(\ref{eq10}) indeed has the minimum Hilbert-Schmidt norm, we need to calculate the instantaneous energy variance $[\Delta H(t)]_{\rho_i(t)}^2=\operatorname{Tr}[\rho_i(t) H^2(t)]-(\operatorname{Tr}[\rho_i(t) H(t)])^2,$ where $\rho_i(t)=\ketz{\tilde{\psi}_i(t)}\braz{\tilde{\psi}_i(t)}$ and $\ketz{\tilde{\psi}_i}$ is the instantaneous eigenstate of $\rho(t)$, assuming that $\rho_i(t)$ evolves unitarily. By making use of the fact that $\braketz{\tilde{\psi}_i(t)}{\tilde{\psi}_j(t)}=\delta_{ij}$ and $\braketz{\tilde{\psi}_i(t)}{\partial_t \tilde{\psi}_i(t)}=0$ \cite{hu2022quantum}, we obtain
\begin{equation}
	\begin{aligned}
		\begin{aligned}{ [\Delta H(t)]}^{2}_{\rho_i(t)} & =\operatorname{Tr} [\rho_{i}(t) H^{2}(t) ]- (\operatorname{Tr} [\rho_{i}(t) H(t) ] )^{2} \\ 
		&=\sum_{ j k l} \langle\tilde{\psi}_{l}(t)\underbrace{|\tilde{\psi}_{i}(t) \rangle \langle\tilde{\psi}_{i}(t)|}_{\rho_i(t)}\underbrace{\partial_{t} \tilde{\psi}_{j}(t) \rangle \langle\tilde{\psi}_{j}(t)|\tilde{\psi}_{k}(t) \rangle \langle\partial_{t} \tilde{\psi}_{k}(t)}_{H^2(t)}|\tilde{\psi}_{l}(t) \rangle+(\underbrace{ \langle\tilde{\psi}_{i}(t) | \partial_{t} \tilde{\psi}_{i}(t) \rangle}_{=0 \text { for all } i})^{2}\\
		&=\sum_k|\braketz{\tilde{\psi}_i(t)}{\partial_t \tilde{\psi}_k(t)}|^2,\\
\end{aligned}
	\end{aligned}
\end{equation}
Since the instantaneous energy variance $[\Delta H(t)]_{\rho_i(t)}^2$ for a given trajectory $\rho_i(t)$ is a fixed value independent of the explicit form of $H(t)$ (provided that $H(t)$ solves $\rho_i(t)$) \cite{hu2023describing}, a minimized $(\operatorname{Tr}[\rho_i(t) H(t)])^2=0$ implies a minimized $\operatorname{Tr}[\rho_i(t) H^2(t)]$.
By
\begin{equation}
\operatorname{Tr}[H^2(t)]=\operatorname {Tr} {\left [\sum_i \rho_i (t) H^2(t)\right]}=\sum_i\operatorname {Tr} {\left [ \rho_i (t) H^2(t)\right]},
\end{equation}
we conclude that for a given $\rho(t)$, if $H(t)$ solves every instantaneous eigenstate of $\rho(t)$ and $\operatorname {Tr} {  [ \rho_i (t) H^2(t) ]}$ is minimized for all $i$, then $\operatorname{Tr}[H^2(t)]$ is also minimized.
\\\\
 \setcounter{equation}{0}
\setcounter{figure}{0}
\setcounter{table}{0}
\renewcommand{\theequation}{D\arabic{equation}}
\renewcommand{\thefigure}{D\arabic{figure}}
\section{Generalization to discrete dynamics\label{appendix:d}}
In the following, we will derive Eq.\,(\ref{eq12}) in the main text.
Denote the eigenvalues of $\rho_D(t)$ by $p_1(t), \cdots, p_d(t)$ and the eigenvalues of $\rho_D(t+s)$ by $p_1(t)+f_1(t,s), \cdots, p_d(t)+f_d(t,s)$, where $f_i(t,s)$ denotes the change of $p_i(t)$ during the time interval $[t,t+s]$. By the same method described in the main text, one can obtain a system of equations written in the matrix form
\begin{equation}
\underbrace{\begin{pmatrix}
	p_1&p_d&p_{d-1}&\cdots& p_2\\
	p_2&p_1&p_d &\cdots & p_3\\
	\vdots &&\cdots &&\vdots\\
	p_d &p_{d-1} &p_{d-2}&\cdots&p_1
\end{pmatrix}}_{P}
\underbrace{
\begin{pmatrix}
	q_{0}-1\\
	q_{1}\\
	\vdots\\
	q_{d-1}
\end{pmatrix}}_{\vec{q}}=\underbrace{\begin{pmatrix}
	f_{1}\\
	f_{2}\\
	\vdots\\
	f_{d}
\end{pmatrix}}_{\vec{f}},
\end{equation}
and the expression for the quantum channel for the diagonalized density matrix
\begin{equation}
	\rho_D(t+s)=\sum_{i=0}^{d-1} q_i(t,s)\, \mathcal{U}_i \rho_D(t)\mathcal{U}_i^\dagger.
\end{equation}
Note that here $q_0(t,s) \equiv 1-\sum_{i=1}^{d-1} q_i(t,s)$, different from the definition of $q_0$ in the continuous case as our definition will yield the final result in a nicer form.
By plugging in $\rho_D(t)=\mathbb{U}^\dagger_t \rho(t) \mathbb{U}_t$, $\rho_D(t+s)=\mathbb{U}^\dagger_{t+s} \rho(t+s) \mathbb{U}_{t+s}$ and $\mathbb{U}_{t+s}=U_t(s)\mathbb{U}_{t}$, one obtains
\begin{equation}
	\mathbb{U}^\dagger_{t} U^\dagger_t(s)\rho(t+s)U_t(s)\mathbb{U}_{t}=\sum_{i=0}^{d-1} q_i(t,s) \,\mathcal{U}_i\mathbb{U}^\dagger_t \rho(t) \mathbb{U}_t\mathcal{U}^\dagger_i
\end{equation}
\begin{equation}
	\Rightarrow \rho(t+s)=\sum_{i=0}^{d-1}q_i(t,s)\,\underbrace{U_t(s)\mathbb{U}_{t}\mathcal{U}_i \mathbb{U}_t^\dagger}_{\tilde{\mathcal{U}}_{i;t,s}} \rho(t)\underbrace{\mathbb{U}_t \mathcal{U}^\dagger_i\mathbb{U}^\dagger_{t} U^\dagger_t(s)}_{\tilde{\mathcal{U}}^\dagger_{i;t,s}}
\end{equation}
which recovers Eq.\,(\ref{eq12}). The expression of $U_t(s)$ is non-unique, and a simple choice would be 
\begin{equation}
	U_t(s)=\sum_{i=0}^{d-1} \ket{\psi_i(t+s)}\bra{\psi_i(t)}
\end{equation}
where $\ket{\psi_i(t+s)}$ and $\ket{\psi_i(t)}$ are the instantaneously eigenvectors of $\rho(t+s)$ and $\rho(t)$, respectively.

 \setcounter{equation}{0}
\setcounter{figure}{0}
\setcounter{table}{0}
\renewcommand{\theequation}{E\arabic{equation}}
\renewcommand{\thefigure}{E\arabic{figure}}
\section{Generalization to the infinite-dimensional system \label{appendix:e}}
In the following, we will generalize the results in the main text to the cases of countably infinite-dimensional Hilbert space. The key is to generalize Eq.\,(\ref{eq7}), i.e., find an infinite-dimensional matrix $P$ and the shift unitary operator $\mathcal{U}_i$ that plays a similar role of real Weyl operator in finite-dimensional case. It turns out that instead of a circulant matrix, the matrix $P$ can be defined as a lower triangular Toeplitz matrix in the infinite-dimensional case, such that
\begin{equation}
	\underbrace{\begin{pmatrix}
		p_1&p_{0}&p_{-1}&p_{-2}&\cdots\\
		p_2&p_{1}&p_{0}&p_{-1}&\cdots\\
		p_3&p_{2}&p_{1}&p_{0}&\cdots\\
		p_4&p_{3}&p_{2}&p_{1}&\cdots\\
		\vdots&\vdots&\vdots&\vdots&\ddots\\
	\end{pmatrix}}_{P} \underbrace{\begin{pmatrix}
		-q_0\\
		q_1\\
		q_2\\
		q_3\\
		\vdots
	\end{pmatrix}}_{\vec{q}}=\underbrace{\begin{pmatrix}
		f_1\\
		f_2\\
		f_3\\
		f_4\\
		\vdots
	\end{pmatrix}}_{\vec{f}},
\end{equation}
where $p_0=p_{-1}=p_{-2}=\dots=0,$ and $p_1,p_2,\dots$, are the eigenvalues of the density matrix $\rho(t)$. Since any infinite-dimensional triangular matrix is invertible if and only if all entries on the diagonal are nonzero, and since we can always rearrange the eigenvalues of the density matrix such that $p_1$ is nonzero, the matrix $P$ can always be made invertible in the infinite-dimensional case. Moreover, since
\begin{equation}
f_n=\sum_{j=1}^\infty P_{nj}[\vec{q}]_j=\sum_{j=1}^\infty p_{n-j}[\vec{q}]_j,
\end{equation}
and
\begin{equation}
	\vec{1}\cdot\vec{f}=\sum_{n=1}^\infty f_n
\end{equation}
we have
\begin{equation}
\sum_{n=1}^\infty f_n=\vec{1}\cdot\vec{f}=\vec{1}\cdot P \cdot \vec{q}=(\sum_{i=1}^\infty p_i, \sum_{i=1}^\infty p_{i-1}, \sum_{i=1}^\infty p_{i-2},\cdots) \vec{q}=(1,1,1\dots)\vec{q}=\sum_{n=1}^\infty[\vec{q}]_n,
\end{equation}
meaning that $\vec{q}$ conserves the probability (i.e., $\sum_{j=1}^\infty [\vec{q}]_j=0$) if and only if $\vec{f}$ conserves the trace (i.e., $\sum_{n=1}^\infty f_n=0$). The corresponding shift operator $\mathcal{U}_i$ associated with $q_i$ is defined as
\begin{equation}
	\mathcal{U}_i=\sum_{m=0}^\infty\ket{m+i}\bra{m}, \quad i\in\mathbb{N}.
\end{equation}
One can then write down the infinite-dimensional master equation Eq.\,(\ref{eq17}).
 \setcounter{equation}{0}
\setcounter{figure}{0}
\setcounter{table}{0}
\renewcommand{\theequation}{F\arabic{equation}}
\renewcommand{\thefigure}{F\arabic{figure}}
\section{A more general result under the probabilistic unitary master equation\label{appendix:f}}
While in Example II, we have shown a special case where the decay of a two-level atom is described by the probabilistic unitary master equation (Eq.\,(24)), in the following, we will show its general expression, assuming an arbitrary initial state. Since the probabilistic unitary master equation is state-dependent, we need to know the explicit expression of $\rho(t)$ to write the master equation down. Assuming $H=\frac{1}{2} \omega \sigma_z$ and 
\begin{equation}
\rho(t)=\begin{pmatrix}
	p_{11}(t) & p_{12}(t)\\
	p_{21}(t) & p_{22}(t)
\end{pmatrix},
\end{equation}
we can solve Eq.\,(23) to obtain

\begin{equation}
\begin{aligned}
	p_{11}(t)&=p_{11}(0)e^{-\Gamma t}\\
	p_{12}(t)&=p_{12}(0)e^{-\frac{\Gamma t}{2}-i\omega t}\\
	p_{21}(t)&=p_{21}(0)e^{-\frac{\Gamma t}{2}+i\omega t}\\
	p_{22}(t)&=1-p_{11}(0)e^{-\Gamma t}.
\end{aligned}
\end{equation}
Since the density matrix $\rho(t)$ has eigenvalues
\begin{equation}
\begin{aligned}
\lambda_1(t)&=\left(\frac{1}{2}\left(1-\sqrt{4 p_{11}^{2}(t)-4 p_{11}(t)+4 |p_{12}(t)|^2+1}\right)\right.\\
\lambda_2(t)&=\left(\frac{1}{2}\left(1+\sqrt{4 p_{11}^{2}(t)-4 p_{11}(t)+4 |p_{12}(t)|^2+1}\right)\right.,
\end{aligned}
\end{equation}
and normalized eigenvectors
\begin{equation} \label{eq:f4}
	\begin{aligned}
		\ket{\psi_1(t)}&=\left(\frac{p_{11}(t)-\lambda_{2}(t)}{p_{21} \sqrt{\frac{\left(\lambda_{2}(t)-p_{11}(t)\right)^{2}}{\left|p_{12}(t)\right|^{2}}+1}}, \frac{1}{ \sqrt{\frac{\left(\lambda_{2}(t)-p_{11}(t)\right)^{2}}{\left|p_{12}(t)\right|^{2}}+1}} \right)^T\\
		\ket{\psi_2(t)}&=\left(\frac{p_{11}(t)-\lambda_{1}(t)}{p_{21} \sqrt{\frac{\left(\lambda_{1}(t)-p_{11}(t)\right)^{2}}{\left|p_{12}(t)\right|^{2}}+1}}, \frac{1}{ \sqrt{\frac{\left(\lambda_{1}(t)-p_{11}(t)\right)^{2}}{\left|p_{12}(t)\right|^{2}}+1}} \right)^T,
	\end{aligned}
\end{equation}
it can be diagonalized as
$$\rho(t)=\mathbb{U}_t \begin{pmatrix}
	\lambda_1(t)&0\\
	0&\lambda_2(t)
\end{pmatrix} \mathbb{U}^\dagger_t,$$
where $\mathbb{U}_t=(\ket{\psi_1(t)}, \ket{\psi_2(t)})$. To solve for $q_1(t)$, we require
\begin{equation}
\begin{pmatrix}
	\lambda_1(t) &\lambda_2(t)\\
	\lambda_2(t) & \lambda_1(t)
\end{pmatrix}\begin{pmatrix}
	-q_0(t)\\
	q_1(t)
\end{pmatrix}=\begin{pmatrix}
	\dot{\lambda}_1(t)\\
\dot{\lambda}_2(t)
\end{pmatrix},
\end{equation}
which gives 
\begin{equation}
	 q_1(t)=q_0(t)=\frac{\dot{\lambda}_2(t) \lambda_1(t)-\dot{\lambda}_1(t) \lambda_2(t)}{\lambda_2^2(t)-\lambda_1^2(t)}=\frac{\dot{\lambda}_1(t)}{1-2\lambda_1(t)},
\end{equation}
where we have used $\lambda_1(t)+\lambda_2(t)=1$ and $\dot{\lambda}_1(t)+\dot{\lambda}_2(t)=0$.

Therefore, the general probabilistic unitary master equation for the decay of a two-level system with $H_S=\frac{1}{2}\omega \sigma_z$ is given by
\begin{equation} \label{eqs21}
	\dot{\rho}(t)=-i[H(t),\rho(t)]+\frac{\dot{\lambda}_1(t)}{1-2\lambda_1(t)}\tilde{\mathcal{U}}_{t}\rho(t)\tilde{\mathcal{U}}^\dagger_t,
\end{equation}
where
\begin{equation} \label{eqs22}
	\tilde{\mathcal{U}}_t=\mathbb{U}_t\sigma_x \mathbb{U}^\dagger_t,
\end{equation}
and $\mathbb{U}_t=(\ket{\psi_1(t)}, \ket{\psi_2(t)})$. $\ket{\psi_i(t)}$, defined by Eq.\,(\ref{eq:f4}), are the instantaneous eigenvectors of $\rho(t)$. The (non-optimal) Hamiltonian $H(t)$ can given by
\begin{equation} \label{eqs23}
H(t)=i\sum_{i=1}^2 \ket{\partial _t\psi_i(t)}\bra{\psi_i(t)}.
\end{equation}
\end{widetext}
\bibliographystyle{ieeetr}
\bibliography{apssamp3}

\begin{thebibliography}{10}

\bibitem{landau1927dampfungsproblem}
L.~Landau, ``Das d{\"a}mpfungsproblem in der wellenmechanik,'' {\em Zeitschrift f{\"u}r Physik}, vol.~45, no.~5-6, pp.~430--441, 1927.

\bibitem{lamb1964theory}
W.~E. Lamb~Jr, ``Theory of an optical maser,'' {\em Physical Review}, vol.~134, no.~6A, p.~A1429, 1964.

\bibitem{redfield1965theory}
A.~Redfield, ``The theory of relaxation processes,'' in {\em Advances in Magnetic and Optical Resonance}, vol.~1, pp.~1--32, Elsevier, 1965.

\bibitem{davies1974markovian}
E.~B. Davies, ``Markovian master equations,'' {\em Communications in mathematical Physics}, vol.~39, pp.~91--110, 1974.

\bibitem{KOSSAKOWSKI1972247}
A.~Kossakowski, ``On quantum statistical mechanics of non-hamiltonian systems,'' {\em Reports on Mathematical Physics}, vol.~3, no.~4, pp.~247--274, 1972.

\bibitem{gorini1976completely}
V.~Gorini, A.~Kossakowski, and E.~C.~G. Sudarshan, ``Completely positive dynamical semigroups of n-level systems,'' {\em Journal of Mathematical Physics}, vol.~17, no.~5, pp.~821--825, 1976.

\bibitem{lindblad1976generators}
G.~Lindblad, ``On the generators of quantum dynamical semigroups,'' {\em Communications in Mathematical Physics}, vol.~48, pp.~119--130, 1976.

\bibitem{breuer2002theory}
H.-P. Breuer, F.~Petruccione, {\em et~al.}, {\em The theory of open quantum systems}.
\newblock Oxford University Press on Demand, 2002.

\bibitem{nakajima1958quantum}
S.~Nakajima, ``On quantum theory of transport phenomena: steady diffusion,'' {\em Progress of Theoretical Physics}, vol.~20, no.~6, pp.~948--959, 1958.

\bibitem{zwanzig1960ensemble}
R.~Zwanzig, ``Ensemble method in the theory of irreversibility,'' {\em The Journal of Chemical Physics}, vol.~33, no.~5, pp.~1338--1341, 1960.

\bibitem{kubo1963stochastic}
R.~Kubo, ``Stochastic liouville equations,'' {\em Journal of Mathematical Physics}, vol.~4, no.~2, pp.~174--183, 1963.

\bibitem{chaturvedi1979time}
S.~Chaturvedi and F.~Shibata, ``Time-convolutionless projection operator formalism for elimination of fast variables. applications to brownian motion,'' {\em Zeitschrift f{\"u}r Physik B Condensed Matter}, vol.~35, no.~3, pp.~297--308, 1979.

\bibitem{PhysRevA.74.053815}
A.~A. Budini, ``Lindblad rate equations,'' {\em Phys. Rev. A}, vol.~74, p.~053815, Nov 2006.

\bibitem{PhysRevA.75.022103}
H.-P. Breuer, ``Non-markovian generalization of the lindblad theory of open quantum systems,'' {\em Phys. Rev. A}, vol.~75, p.~022103, Feb 2007.

\bibitem{rybar2012simulation}
T.~Ryb{\'a}r, S.~N. Filippov, M.~Ziman, and V.~Bu{\v{z}}ek, ``Simulation of indivisible qubit channels in collision models,'' {\em Journal of Physics B: Atomic, Molecular and Optical Physics}, vol.~45, no.~15, p.~154006, 2012.

\bibitem{ciccarello2013quantum}
F.~Ciccarello and V.~Giovannetti, ``A quantum non-markovian collision model: incoherent swap case,'' {\em Physica Scripta}, vol.~2013, no.~T153, p.~014010, 2013.

\bibitem{PhysRevA.87.040103}
F.~Ciccarello, G.~M. Palma, and V.~Giovannetti, ``Collision-model-based approach to non-markovian quantum dynamics,'' {\em Phys. Rev. A}, vol.~87, p.~040103, Apr 2013.

\bibitem{imamog1994stochastic}
A.~Imamog {\em et~al.}, ``Stochastic wave-function approach to non-markovian systems,'' {\em Physical Review A}, vol.~50, no.~5, p.~3650, 1994.

\bibitem{diosi1997non}
L.~Di{\'o}si and W.~T. Strunz, ``The non-markovian stochastic schr{\"o}dinger equation for open systems,'' {\em Physics Letters A}, vol.~235, no.~6, pp.~569--573, 1997.

\bibitem{PhysRevA.58.1699}
L.~Di\'osi, N.~Gisin, and W.~T. Strunz, ``Non-markovian quantum state diffusion,'' {\em Phys. Rev. A}, vol.~58, pp.~1699--1712, Sep 1998.

\bibitem{PhysRevLett.124.190402}
A.~Smirne, M.~Caiaffa, and J.~Piilo, ``Rate operator unraveling for open quantum system dynamics,'' {\em Phys. Rev. Lett.}, vol.~124, p.~190402, May 2020.

\bibitem{RevModPhys.89.015001}
I.~de~Vega and D.~Alonso, ``Dynamics of non-markovian open quantum systems,'' {\em Rev. Mod. Phys.}, vol.~89, p.~015001, Jan 2017.

\bibitem{bertlmann2008bloch}
R.~A. Bertlmann and P.~Krammer, ``Bloch vectors for qudits,'' {\em Journal of Physics A: Mathematical and Theoretical}, vol.~41, no.~23, p.~235303, 2008.

\bibitem{peetz2023simulation}
J.~Peetz, S.~E. Smart, S.~Tserkis, and P.~Narang, ``Simulation of open quantum systems via low-depth convex unitary evolutions,'' {\em arXiv preprint arXiv:2307.14325}, 2023.

\bibitem{feynman1984negative}
R.~P. Feynman, ``Negative probability,'' tech. rep., PRE-27827, 1984.

\bibitem{nielsen2002quantum}
M.~A. Nielsen and I.~Chuang, ``Quantum computation and quantum information,'' 2002.

\bibitem{alipour2020shortcuts}
S.~Alipour, A.~Chenu, A.~T. Rezakhani, and A.~del Campo, ``Shortcuts to adiabaticity in driven open quantum systems: Balanced gain and loss and non-markovian evolution,'' {\em Quantum}, vol.~4, p.~336, 2020.

\bibitem{hu2020quantum}
Z.~Hu, R.~Xia, and S.~Kais, ``A quantum algorithm for evolving open quantum dynamics on quantum computing devices,'' {\em Scientific reports}, vol.~10, no.~1, p.~3301, 2020.

\bibitem{PhysRevLett.127.270503}
A.~W. Schlimgen, K.~Head-Marsden, L.~M. Sager, P.~Narang, and D.~A. Mazziotti, ``Quantum simulation of open quantum systems using a unitary decomposition of operators,'' {\em Phys. Rev. Lett.}, vol.~127, p.~270503, Dec 2021.

\bibitem{suri2023two}
N.~Suri, J.~Barreto, S.~Hadfield, N.~Wiebe, F.~Wudarski, and J.~Marshall, ``Two-unitary decomposition algorithm and open quantum system simulation,'' {\em Quantum}, vol.~7, p.~1002, 2023.

\bibitem{kra2012circulant}
I.~Kra and S.~R. Simanca, ``On circulant matrices,'' {\em Notices of the AMS}, vol.~59, no.~3, pp.~368--377, 2012.

\bibitem{hu2022quantum}
L.~Hu and A.~N. Jordan, ``Quantum state driving along arbitrary trajectories,'' {\em arXiv preprint arXiv:2211.02457}, 2022.

\bibitem{PhysRevLett.73.1060}
P.~Pechukas, ``Reduced dynamics need not be completely positive,'' {\em Phys. Rev. Lett.}, vol.~73, pp.~1060--1062, Aug 1994.

\bibitem{shaji2005s}
A.~Shaji and E.~C.~G. Sudarshan, ``Who's afraid of not completely positive maps?,'' {\em Physics Letters A}, vol.~341, no.~1-4, pp.~48--54, 2005.

\bibitem{audenaert2008random}
K.~M. Audenaert and S.~Scheel, ``On random unitary channels,'' {\em New Journal of Physics}, vol.~10, no.~2, p.~023011, 2008.

\bibitem{watrous2018theory}
J.~Watrous, {\em The theory of quantum information}.
\newblock Cambridge university press, 2018.

\bibitem{lee2020detecting}
C.~D.-Y. Lee and J.~Watrous, ``Detecting mixed-unitary quantum channels is np-hard,'' {\em Quantum}, vol.~4, p.~253, 2020.

\bibitem{jaynes1963comparison}
E.~T. Jaynes and F.~W. Cummings, ``Comparison of quantum and semiclassical radiation theories with application to the beam maser,'' {\em Proceedings of the IEEE}, vol.~51, no.~1, pp.~89--109, 1963.

\bibitem{berman2011principles}
P.~R. Berman and V.~S. Malinovsky, {\em Principles of laser spectroscopy and quantum optics}.
\newblock Princeton University Press, 2011.

\bibitem{chruscinski2010non}
D.~Chru{\'s}ci{\'n}ski and A.~Kossakowski, ``Non-markovian quantum dynamics: local versus nonlocal,'' {\em Physical review letters}, vol.~104, no.~7, p.~070406, 2010.

\bibitem{yu2004finite}
T.~Yu and J.~Eberly, ``Finite-time disentanglement via spontaneous emission,'' {\em Physical Review Letters}, vol.~93, no.~14, p.~140404, 2004.

\bibitem{yu2006sudden}
T.~Yu and J.~Eberly, ``Sudden death of entanglement: classical noise effects,'' {\em Optics Communications}, vol.~264, no.~2, pp.~393--397, 2006.

\bibitem{hu2023signs}
L.~Hu and A.~N. Jordan, ``Signs of the rates in the lindblad master equations can always be arbitrarily determined,'' {\em arXiv preprint arXiv:2310.17881}, 2023.

\bibitem{hu2023describing}
L.~Hu and A.~N. Jordan, ``Describing the wave function collapse process with a state-dependent hamiltonian,'' {\em arXiv preprint arXiv:2301.09274}, 2023.

\end{thebibliography}

\end{document}